\def \SAIT #1 #2 {{\em Mem.\ Soc.\ Astron.\ It.\/} {\bf #1}, #2}
\def \MESS #1 #2 {{\em The Messenger\/} {\bf #1}, #2}
\def \ASTRNACH #1 #2 {{\em Astron. Nach.\/} {\bf #1}, #2}
\def \AAP #1 #2 {{\em Astron. Astrophys.\/} {\bf #1}, #2}
\def \AAL #1 #2 {{\em Astron. Astrophys. Lett.\/} {\bf #1}, L#2}
\def \AAR #1 #2 {{\em Astron. Astrophys. Rev.\/} {\bf #1}, #2}
\def \AAS #1 #2 {{\em Astron. Astrophys. Suppl. Ser.\/} {\bf #1}, #2}
\def \AJ #1 #2 {{\em Astron. J.\/} {\bf #1}, #2}
\def \ANNREV #1 #2 {{\em Ann. Rev. Astron. Astrophys.\/} {\bf #1}, #2}
\def \APJ #1 #2 {{\em Astrophys. J.\/} {\bf #1}, #2}
\def \APJL #1 #2 {{\em Astrophys. J. Lett.\/} {\bf #1}, L#2}
\def \APJS #1 #2 {{\em Astrophys. J. Suppl.\/} {\bf #1}, #2}
\def \APSS #1 #2 {{\em Astrophys. Space Sci.\/} {\bf #1}, #2}
\def \ASR #1 #2 {{\em Adv. Space Res.\/} {\bf #1}, #2}
\def \BAIC #1 #2 {{\em Bull. Astron. Inst. Czechosl.\/} {\bf #1}, #2}
\def \JSQRT #1 #2 {{\em J. Quant. Spectrosc. Radiat. Transfer\/} {\bf #1}, #2}
\def \MN #1 #2 {{\em Mon. Not. R. Astr. Soc.\/} {\bf #1}, #2}
\def \MEM #1 #2 {{\em Mem. R. Astr. Soc.\/} {\bf #1}, #2}
\def \PLR #1 #2 {{\em Phys. Lett. Rev.\/} {\bf #1}, #2}
\def \PASJ #1 #2 {{\em Publ. Astron. Soc. Japan\/} {\bf #1}, #2}
\def \PASP #1 #2 {{\em Publ. Astr. Soc. Pacific\/} {\bf #1}, #2}
\def \NAT #1 #2 {{\em Nature\/} {\bf #1}, #2}

\documentstyle{memsait}
\input epsf.sty
\begin{opening}
\title{ON THE CRAB PROPER MOTION.} 
\author{PATRIZIA A. CARAVEO$^{1,2}$ and   ROBERTO MIGNANI$^3$}
\institute{$^1$Istituto di Fisica Cosmica del CNR , Milano, Italy\\
$^2$Istituto Astronomico, Roma, Italy\\
$^3$STECF-ESO, Garching, Germany}
\date{} 
\end{opening}

\begin{document}

\oddpagefooter{}{}{} 
\evenpagefooter{}{}{} 
\ 
\bigskip

\begin{abstract}
Owing to the
dramatic evolution of telescopes as well as optical detectors in the last 20
yrs, we are now able to measure anew
the proper motion of the Crab pulsar, after the classical result
of Wyckoff and  Murray  (1977) in a time span 40 times shorter.
The proper motion is aligned with the axis of symmetry of the inner Crab 
nebula and, presumably, with the pulsar spin axis.
\end{abstract}

\section{Introduction}

Isolated Neutron  Stars are fast  moving  objects (e.g. Caraveo, 1993,
Lyne and Lorimer, 1993; Lorimer, 1998), and  the Crab is no exception.
The measure of the proper motion of Baade's  star (later recognized to
be the  optical counterpart of  the Crab pulsar) was attempted several
times (e.g.  Trimble,  1968), yielding  vastly different values.  This
prompted  Minkowski (1970) to conclude  that  the proper motion of the
star  was not reliably  measured.   The situation changed  few years
later, when Wyckoff and Murray (1977) obtained a new value of the Crab
proper  motion which allowed to reconcile  the  pulsar birthplace with
the center of the  nebula, i.e.  the  filaments' divergent  point. The
relative proper  motion measured  by  Wyckoff and Murray  amounts to a
total  yearly  displacement of $15   \pm 3 ~mas$,  corresponding  to a
transverse velocity of  123 km/s for a  pulsar  distance of 2 kpc.  \\
Recently, Hester et al  (1995) have drawn a  convincing picture of the
central  part  of the Crab  Nebula   "symmetrical about the (presumed)
rotation  axis of the pulsar" by  associating the ROSAT/HRI picture of
the pulsar and its surroundings  with HST/WFPC2 images of the remnant.
However, they failed to note that  the position angle  of such an axis
of symmetry is fully compatible with  the position angle of the proper
motion  vector measured by Wyckoff  and Murray (1977).  Since the link
between  pulsars' proper  motions and their  spin  axes has been the
subject of  many inconclusive studies, we have sought an independent
measurement of the pulsar proper motion. 
 
\section{The Method and the Data}

In order to measure the tiny angular  displacement of the Crab pulsar,
we need high resolution images taken at different epochs, like the ones
available in   HST  Public Archive. Browsing through the HST archive,
one finds for the Crab about 60 entries, 30 of which are images taken with
the different instruments used troughout the mission: FOC, WFPC,
STIS and NICMOS.  Since   the WFPC2 was the
most commonly used, both as Wide Field and as Planetary Camera, we focussed
on it, selecting images   taken through the  same filter.
The 547M medium bandpass ($\lambda= 5454\AA; 
\Delta \lambda=486.6\AA$) turned out to  be the most popular, totalling 12
images, with the Crab pulsar positioned either in
one of  the  three Wide Field  Camera  chips or,  more  often, in the
Planetary  Camera.   The {\it 547M} data  set  has  been retrieved and
after cosmic ray  cleaning, all images have been
inspected to define  a suitable set  of reference stars. 
When doing astrometric studies the presence of good reference
stars is very  important: an outstanding  image  without  at least  4
reference  objects, 
is of no use for our purposes.  
While finding reference stars for the Wide Field images is not a problem, 
the smaller field of view of the Planetary Camera rarely 
contained enough reference objects. Indeed, only the PC observation shown in 
Figure 1 meets our requirements.

\begin{figure}
\epsfysize=8cm 
\hspace{3.5cm}
\epsfbox{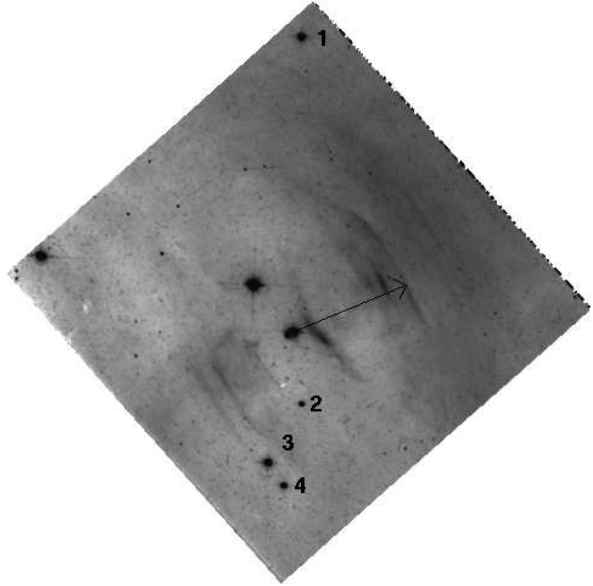}  
\caption[h]{2000 sec PC image of the Crab Nebula taken in Aug 
1995, through the F547M filter. North to the top, East to the left.
The labels mark the stars used for relative astrometry. The arrow shows the
Crab pulsar proper motion direction.}
\end{figure}

The need for reference stars limits the data-set 
to just three images,
taken   
in march 94  (WFC-chip\#2), august 95 (PC)
and january 96  (WFC-chip\#2).    
The images  have   been
superimposed following the  rotate-shift procedure outlined in Caraveo
et al (1996).   First, the frames  have been  corrected for the  WFPC2
geometrical   distorsion  (Holtzman  et    al,  1995) and    the scale
transformation from the PC  (0''.045/px)  to the WFC (0''.1/px)  scale
was applied.  Next,  the frames have  been aligned in right  ascension
and declination  according   to  their  roll angles   and   the "best"
positions of the Crab pulsar, as well  as of the reference stars, have
been  computed by 2-D gaussian fitting  of their profiles.  Particular
care was used for  the pulsar itself in  order  to make sure  that the
object's centroid is not affected by the emission knot observed $\sim$
0.7 arcsec  to the SE.  A positional  accuracy ranging from 0.02 px to
0.03  px was  achieved  for the pulsar   as well as  for the reference
stars.  Finally, we used the common reference stars (1 to 4 in Fig. 1)
to compute the  linear shifts needed to  overlay  the different frames
onto the march 94 image , which was used as a reference. 

\section{Results}

With three images accurately superimposed we can compare the positions
obtained for the  Crab pulsars over 1.9  yrs.  While the positions  of
the  reference  stars at   the three  different epochs  are  virtually
unchanged, the pulsar is clearly affected by a proper motion to NW.  A
linear fit to  the  ${\alpha}$ and ${\delta}$ displacements  (shown in
Fig.  2a,b)  yields the Crab  proper  motion relative to the reference
stars. This turns out to be 

$$\mu_{\alpha}=-17   \pm  3     ~mas~yr^{-1},   \mu_{\delta}=  7   \pm
3~mas~yr^{-1}$$ 

\noindent

corresponding  to  an overall  annual  displacement  $\mu  = 18  \pm 3
~mas~yr^{-1}$  in  the plane  of  the sky,  with  a position  angle of
$292^{\circ} \pm 10^{\circ}$.  This vector  is also plotted in  figure
1. \\ Our result is to be compared with the value of   

$$\mu_{\alpha}=-13 \pm 2 ~mas~yr^{-1}, \mu_{\delta}= 7 \pm 3~mas~yr^{-1}$$

obtained by Wyckoff and Murray (1977).

\begin{figure}
\epsfysize=5.5cm 
\hspace{0.cm}\epsfbox{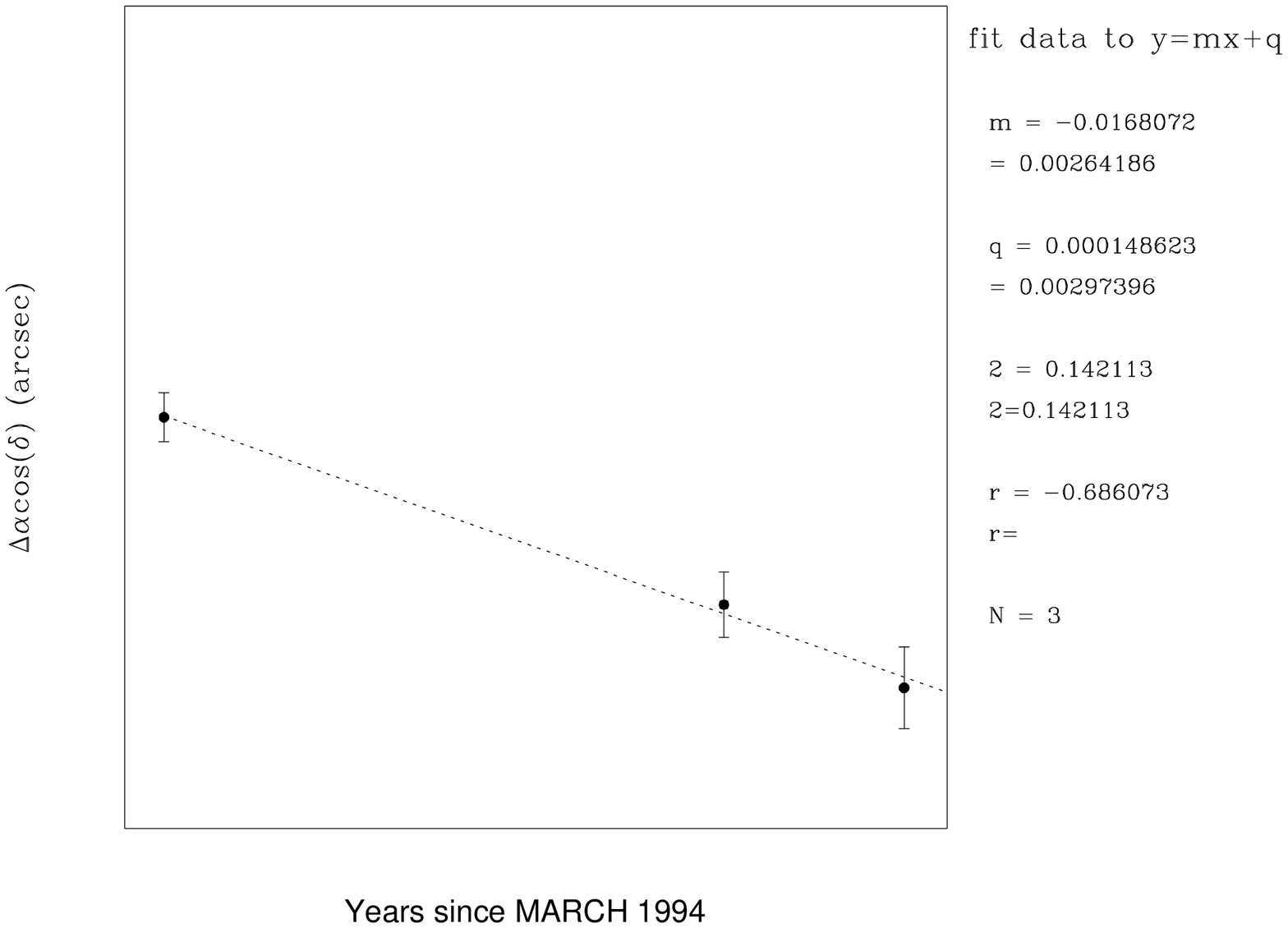} 
\epsfysize=5.5cm 
\hspace{1.5cm}\epsfbox{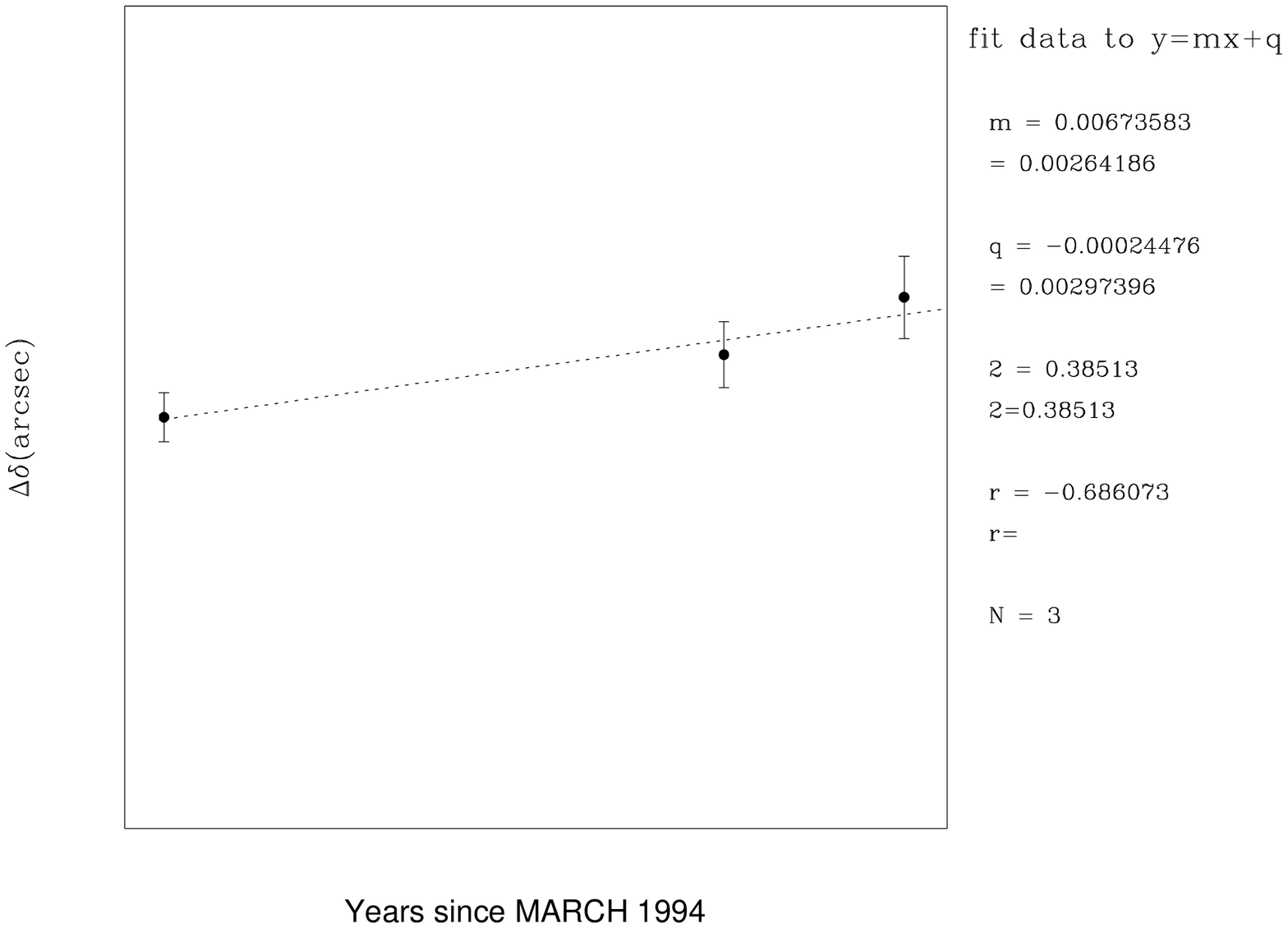}
\caption[h]{Linear fit to the Crab pulsar displacements in RA an DEC, 
respectively.}
\end{figure}

\section{Conclusions}

According to Hester et al (1995), the position angle  of the axis of symmetry
of the inner Crab Nebula is $\sim 115^{\circ}$, or, rather, $\sim 295^{\circ}$,
to take into account an $180^{\circ}$ offset.    
This value is to be compared to
our result $\sim  292^{\circ}$ or to  that of Wyckoff and Murray ($\sim
298^{\circ}$).  Although a  chance coincidence probability  at a few %
level cannot  be totally dismissed, it is  interesting to speculate on
the implications of such an alignement.  Since a neutron star acquires
its proper motion at  birth, there is no  doubt that the pulsar motion
has been present "ab initio".  However, its energy  content is far too
small   to  account for the   surrounding  structures  and their rapid
evolution. Therefore the link, is  any, between proper motion and axis
of symmetry must go through some  basic characteristics which was also
present when the Crab pulsar was born.\\ Hester  et al (1995) proposed
a scenario associating the  symmetrical appearence of the  Nebula with
the pulsar spin-axis.  Under this hypothesis,  the neutron star motion
would turn  out to   be aligned with   the  spin axis, reflecting   an
asymmetry of the  supernova explosion along  the progenitor spin-axis.
Proper motion  spin-axis alignements    have  been discussed in    the
literature (see e.g. Tademaru,  1977)  but no conclusive evidence  was
found.  If the X-ray jets  do indeed trace  the pulsar spin axis, Crab
would provide the first case of such an alignement.

\end{document}